\documentclass[12pt,a4paper]{article}
\newenvironment{keywords}{Keywords:}{}
\usepackage{fullpage}

\usepackage[utf8]{inputenc}
\usepackage{amsmath,amsfonts}
\usepackage{mathtools} %
\usepackage{stmaryrd} %
\usepackage[usenames,dvipsnames,svgnames,table]{xcolor}
\usepackage{graphicx}
\usepackage{subcaption}
\usepackage[authoryear,round]{natbib}
\usepackage{xfrac}
\usepackage{siunitx}

\usepackage{accents}
\newlength{\dhatheight}

\usepackage[section]{placeins}

\usepackage[colorlinks=false,allbordercolors=white,pdfpagemode=UseOutlines]{hyperref}
\usepackage[noabbrev]{cleveref}

\newcommand{\T}[1]{{\boldsymbol{{#1}}}}
\newcommand{\Op}[1]{{\mathcal{{#1}}}}
\newcommand{\M}[1]{{\mathbf{{#1}}}}
\newcommand{\V}[1]{{\mathbf{{#1}}}}
\newcommand{\parm}{{\mathord{\color{black!33}\bullet}}}
\newcommand{\Vh}[1]{{\hat{\mathbf{{#1}}}}}

\newcommand{\pdiff}[2]{\frac{\partial#1}{\partial#2}}

\newcommand{\dif}{\,\mathrm{d}}

\newcommand{\jump}[1]{\left\llbracket\,#1\,\right\rrbracket}
\newcommand{\avg}[1]{\left\{\!\left\{ #1 \right\}\!\right\}}

\newcommand{\Hdiv}{H\textsuperscript{div}}

\newcommand{\vel}{\T{u}}
\newcommand{\wvel}{\T{w}}
\newcommand{\vtest}{\T{v}}
\newcommand{\pos}{\T{x}}

\newcommand{\element}{K} %

\newcommand{\tess}{\mathcal{T}}
\newcommand{\skel}{\mathcal{S}}

\newcommand{\normal}{\T{n}}

\begin{document}

\title{On exactly incompressible DG FEM pressure splitting schemes for the Navier-Stokes equation}

\author{
  Tormod Landet%
  , %
  Mikael Mortensen
  \\%
  \vspace{6pt}%
  {\em{Department of Mathematics, University of Oslo}}%
  \vspace{-2mm}\\%
  {\em{Moltke Moes vei 35, 0851 Oslo, Norway}}
}
\maketitle

\begin{abstract}
  We compare three iterative pressure correction schemes for solving the Navier-Stokes equations with a focus on exactly divergence free solution with higher order discontinuous Galerkin discretisations. The investigated schemes are the incremental pressure correction scheme on the standard differential form (IPCS-D), the same scheme on algebraic form (IPCS-A), and the semi-implicit method for pressure linked equations (SIMPLE).

  We show algebraically and through numerical examples that the IPCS-A and SIMPLE schemes are exactly mass conserving due to the algebraic pressure correction, while the IPCS-D scheme cannot be exactly divergence free due to the stabilisation terms required in the pressure Poisson equation. The SIMPLE scheme requires a significantly higher number of pressure correction iterations to obtain converged results than the IPCS-A scheme, so for efficient and mass conserving simulation the IPCS-A method is the best option among the three evaluated schemes.

\end{abstract}

\section{Introduction}

The incompressible Navier-Stokes equations contain three major obstacles to constructing an efficient solution procedure: the non-linearity of the convective term, the incompressibility criterion, and the intricate pressure-velocity coupling.
In this work we will focus on how to maintain exact incompressibility when dealing with the lack of a pressure update equation. The non-linearity will simply be handled by introducing an explicitly extrapolated convecting velocity.

Denoting the unknown velocity vector as $\vel$, the explicit convecting velocity as $\wvel$, the pressure as $p$, the fluid density as $\rho$ and the fluid dynamic viscosity as $\mu$, the incompressible Navier-Stokes equations can be written
\begin{align}
  \rho \left( \pdiff{\vel}{t} + (\wvel\cdot\nabla) \vel \right) & = \nabla\cdot\mu\left(\nabla \vel + (\nabla\vel)^T\right) - \nabla p, \label{eq:ns_basis} \\
  \nabla\cdot \vel                                             & = 0, \notag
\end{align}
which, after discretisation, leads to a sparse discrete block matrix problem,
\begin{align}
  \begin{bmatrix}
    \M{A} & \M{B} \\
    \M{C} & \M{0}
  \end{bmatrix}
  \begin{bmatrix}
    \V{u} \\ \V{p}
  \end{bmatrix}
  =
  \begin{bmatrix}
    \V{d} \\ \V{e}
  \end{bmatrix},
  \label{eq:saddle_point_problem}
\end{align}
where the sub-matrix $\M{A}$ is a discrete version of the bilinear operator $\Op{A}$,
\begin{align}
  \Op{A}\,\parm \equiv \frac{\rho}{\Delta t}\gamma_1\parm + \rho(\wvel\cdot\nabla)\parm - \nabla\cdot\mu\left(\nabla \parm + (\nabla\parm)^T\right),
  \label{eq:op_A}
\end{align}
which has been assembled by use of the discontinuous Galerkin finite element (DG FEM) scheme described in \cref{sec:discr}. Similar holds for the matrices $\M{B}$ and $\M{C}$, the discrete versions of the pressure gradient and velocity divergence operators. The unknowns $\V{u}$ and $\V{p}$ (bold, non-italic) are now arrays of unknowns containing the degrees of freedom describing the velocity and pressure fields. The vectors $\V{d}$ and $\V{e}$ contain the assembled linear operators from the momentum and continuity equations respectively. Note that due to boundary conditions the vector $\V{e}$ is not necessarily zero.

There are many methods for solving \cref{eq:saddle_point_problem}. With appropriate boundary and initial conditions the equations are well-formed and can be solved directly on coupled form by an LU decomposition method. Fast parallel sparse LU solvers exist, such as MUMPS, which uses multi\-frontal LU-factorization \citep{MUMPS:1,MUMPS:2}, and SuperLU\_DIST, which uses super\-nodal LU-factorization \citep{superlu05,superlu_ug99,superlu_dist03}. Unfortunately, such direct solution methods only scale up to a relatively small number of parallel processors, but they can deal with the saddle point block matrix problems such as \cref{eq:saddle_point_problem} without any special preconditioning.

Iterative Krylov subspace methods can scale to thousands of processors, but they require preconditioning for efficient and stable solution of saddle point problems \citep{cgmethod52,bi-cg76,gmres86,bi-cgstab92}. Pressure correction methods are a popular family of methods to deal with the saddle point problem in the Navier-Stokes equations by splitting the equation into individual momentum guess and pressure correction steps, which can then be solved efficiently by sparse iterative Krylov solvers. The earliest pressure correction schemes by \citet{Chorin_1968} and \citet{Temam_1969} are still popular, often written on iterative form by a method such as the Incremental Pressure Correction Scheme, IPCS, \citep{ipcs_Goda_1979}. Many other splitting methods exist, such as velocity correction methods \citep{kawahara_velcor_1985,guermond_velocity-correction_2003}, and the more general Schur complement methods \citep{schur_1917,zhang_schur_2005}. The popular PISO \citep{Issa1986} and PIMPLE \citep{Weller_Tabor_Jasak_Fureby_1998} algebraic pressure correction techniques should also be mentioned, and we even implemented them in the Ocellaris flow solver that we have used in this work. But, though they are exactly mass conserving, unfortunately we were unable to make them converge properly with the higher order DG discretisation, so they have been left out of this comparison.

In this paper we will look at the IPCS method on differential form in \cref{sec:ipcs-d} and on algebraic form in \cref{sec:ipcs-a}. The DG SIMPLE method by
\citet{KleinEtAl2013,KleinEtAl2015,KleinEtAl2016} based on the SIMPLE method by \citet{simple_1972} is described in \cref{sec:simple}. The effect of the choice of pressure splitting scheme on the divergence of the resulting velocity field is discussed in \cref{sec:masscons}, before the methods are compared on several benchmark cases in \cref{sec:numexperiments}. 
The DG FEM spatial discretisation and the temporal scheme used in the numerical experiments are presented in \cref{sec:discr}. The findings in this paper are not dependent on the details of the numerical scheme and the results will hold for other methods as long as DG elements are used for the pressure and exact incompressibility can be achieved when the governing equations are solved without pressure-velocity splitting.

\section{The DG FEM discretisation}
\label{sec:discr}

This section contains a brief summary of the DG FEM discretisation of the governing equations \eqref{eq:ns_basis} used to construct the block matrix in \cref{eq:saddle_point_problem}. More details can be found in \citet{landet_slope_2018} which is based on \citet{cockburn_locally_2005}  except for the elliptic term where we have used the symmetric interior penalty (SIP) method by \citet{arnold_interior_1982} and not the local discontinuous Galerkin (LDG) method used by Cockburn et al. The exact details of the presented discretisation are not very important for the focus of this paper. The numerical scheme can be substituted for another numerical scheme as long as (i) the resulting mass matrix is block diagonal, (ii) the pressure is discontinuous and requires continuity-enforcing stabilisation of elliptic operators, and (iii) the velocity field is machine precision divergence free when \cref{eq:saddle_point_problem} is solved on coupled form by a direct solver.

The Navier-Stokes equations from \cref{eq:ns_basis} are first discretised in time by use of values from previous time steps, $(\vel^n, \vel^{n-1})$, along with a suitable choice of time stepping coefficients $(\gamma_1, \gamma_2, \gamma_3)$. The time steps are $t=n\Delta t$ with initial conditions at $t=0$. We have used a second order backwards differencing formulation, BDF2, where the coefficients are $(\gamma_1, \gamma_2, \gamma_3) = (\sfrac{3}{2},\,-2,\,\sfrac{1}{2})$. If the initial value at $t=-\Delta t$ is not provided then coefficients $(1,\,-1,\,0)$ is used for the first time step. The non-linearity of the convective term is handled semi-implicitly by introducing an explicit convecting velocity, $\wvel=2\vel^n-\vel^{n-1}$. The spatial differential equations are then
\begin{align}
  \rho \left( \frac{1}{\Delta t}(\gamma_1\vel + \gamma_2\vel^n + \gamma_3\vel^{n-1}) + (\wvel\cdot\nabla) \vel \right) & = \nabla\cdot\mu\left(\nabla \vel + (\nabla\vel)^T\right) - \nabla p, \label{eq:space_ns} \\
  \nabla\cdot \vel                                                                                                     & = 0. \notag
\end{align}

Let $\tess$ be the set of all cells in the mesh and $\skel$ the set of all facets. Polynomial function spaces of degree $k$ on each cell $\element$ are denoted $P_k(\element)$. These have no continuity at cell boundaries and no inherent boundary conditions. We use calligraphic typeface to denote operators and sets, bold italic for vectors functions and italic for scalar functions. 
\Cref{eq:space_ns} is cast into the following form: find $\vel\in [P_2(\element)]^D$ and $p\in P_{1}(\element)$ such that
\begin{alignat}{4}
\Op{A}(\vel, \vtest; \wvel) + \Op{B}(p, \vtest) &= \Op{D}(\vtest) \quad&&\forall\ \vtest\ &&\in [P_2(\element)]^D, \label{eq:op_nsmom}\\
\Op{C}(\vel, q) &= \Op{E}(q)                                      &&\forall\ q       &&\in P_{1}(\element), \label{eq:op_nssol}
\end{alignat}
in the tessellated domain $\tess$ subject to
\begin{alignat}{2}
\vel &= \vel_D && \quad\text{on the boundary facets,} \ \skel_D\subset\skel. %
\end{alignat}

The discontinuous Galerkin method works by breaking integrals over the whole domain into a sum of integrals over each mesh cell $\element\in\tess$, and defining fluxes of the unknown functions between these cells. The average and jump operators across an internal facet between two cells $\element^+$ and $\element^-$ are defined as 
\begin{align}
\avg{u} &= \frac{1}{2}(u^+ + u^-), \\
\jump{u} &= u^+ - u^-, \\
\jump{\T{u}}_\normal &= \T{u}^+\cdot\normal^+ + \T{u}^-\cdot\normal^-.
\end{align}

The convecting velocity $\wvel$ is \Hdiv-conforming, i.e.\ the flux is continuous, $\jump{\T{w}}_\normal=0$. For exterior facets, let the connected cell be denoted $\element^+$ such that $\normal^+\!\cdot\jump{\vel} = \normal^+\!\cdot\vel^+ = \normal\cdot\vel$. Take $\avg{u}=u$ and otherwise let all $\element^-$ values be zero on the boundary facets.

The momentum equation is discretised using the symmetric interior penalty (SIP) method for the elliptic term \citep{arnold_interior_1982} and otherwise using the fluxes from \citet{cockburn_locally_2005}. The flux of pressure is $\hat{p}=\avg{p}$ and the convective flux $\hat\vel^\wvel$ is a pure upwind flux. After multiplication with $\vtest$ and integration over $\tess$, followed by integration by parts and application of the SIP method to the viscosity; $\Op{A}$ can be found as the bilinear part containing $\vel=\vel^{n+1}$, $\Op{B}$ as the bilinear part containing $p$, and $\Op{D}$ as the linear part of

{\allowdisplaybreaks
\begin{flalign}
&\int_\tess\frac{\rho}{\Delta t}(\gamma_1\vel + \gamma_2\vel^{n} + \gamma_3\vel^{n-1})\vtest\dif\pos
\label{eq:wf_ns_coupled1}
\\ \notag
&\qquad-\ \int_\tess \vel \cdot\nabla\cdot (\rho\vtest\otimes\wvel)\dif\pos
 \ +\ \int_\skel \wvel\cdot\normal^+\,\hat\vel^\wvel\cdot\jump{\rho\vtest}\dif s
\\ \notag
&\qquad+\ \int_\tess \mu\left(\nabla\vel + (\nabla\vel)^T\right):\nabla\vtest\dif\pos
\ +\ \int_{\skel_I}\kappa_\mu\jump{\vel}\cdot\jump{\vtest}\dif s
\\ \notag
&\qquad-\ \int_\skel (\avg{\mu\left(\nabla\vel + (\nabla\vel)^T\right)}\cdot\normal^+)\cdot\jump{\vtest}\dif s
\\ \notag
&\qquad-\ \int_{\skel_I} (\avg{\mu\left(\nabla\vtest + (\nabla\vtest)^T\right)}\cdot\normal^+)\cdot\jump{\vel}\dif s
\\ \notag
&\qquad-\ \int_\tess p\, \nabla\cdot\vtest\dif\pos
 \ +\ \int_\skel\hat{p}\,\normal^+\cdot\jump{\vtest}\dif s
 + \frac{1}{2} \int_\tess (\nabla\cdot\wvel) \vel\cdot\vtest \dif\pos = 0
.
\end{flalign}}

On Dirichlet boundaries $\hat\vel^\wvel$ is the upwind value. Depending on flow direction this is either $\vel$ or $\vel_D$. On all boundaries take $\hat{p}=p$. The SIP penalty parameter is selected based on \citet{epshteyn_estimation_2007,shahbazi_2007},
\begin{align}
  \kappa_\mu = 3\,\frac{\mu_\mathrm{max}^2}{\mu_\mathrm{min}}\,k(k+1)\,\max_K\left(\frac{S_K}{V_K}\right),
  \label{eq:penalty_param}
\end{align}
where the $k$ is the order of the approximating polynomials, $S_K$ is the cell surface area and $V_K$ the cell volume. The penalty parameter is doubled on external facets. 

Since the IPCS-D scheme is not exactly divergence free, the standard skew symmetric term \citep{gresho_cfd_issues_1991} is included in the weak form used for all the methods. The term is the last integral in \cref{eq:wf_ns_coupled1},
\begin{align}
  \frac{1}{2} \int_\tess (\nabla\cdot\wvel) \vel\cdot\vtest \dif\pos.
  \label{eq:wf_ns_skewsymm}
\end{align}

The continuity equation is multiplied by $q$ and integrated over $\tess$. The flux of velocity is here $\hat{\vel}^p=\avg{\vel}$. After integration by parts $\Op{C}(\vel,q)$ and $\Op{E}(q)$ can be found from
\begin{align}
&\int_\skel \hat{\vel}^p\cdot\normal^+\jump{q} \dif s
 \ -\ \int_\tess \vel\cdot\nabla q\dif\pos
 \ =\ 0.
\label{eq:wf_ns_coupled2}
\end{align}
The non-zero $\Op{E}(q)$ results from using the boundary conditions in the flux, $\hat\vel^p=\vel_D$ on Dirichlet boundaries.

After solving \cref{eq:op_nsmom,eq:op_nssol} by one of the presented pressure splitting schemes, the resulting velocity field is projected into a space where it is exactly divergence free and where the normal velocities across internal facets are continuous. This projection from weak to strong incompressibility requires only that the discrete version of the weak incompressibility criterion on the form shown in \cref{eq:wf_ns_coupled2} is satisfied exactly. The projection is described in \citet{cockburn_locally_2005}. It is not necessary to perform the projection inside the pressure iteration loops, it is run once as a post processing operation at the end of each time step.

\subsection{Differential Poisson equation for the pressure}
\label{sec:ipcs-d-elliptic-stab}

In the IPCS-D method, which will be described in \cref{sec:ipcs-d}, a Poisson equation must be solved for the pressure,
\begin{align}
  \nabla\cdot\left(\frac{1}{\rho}\nabla p\right) = f.
\end{align}
This equation is discretised with the same SIP penalty method  used for the viscous term in the momentum equation  and the same penalty parameter as in \cref{eq:penalty_param} is used, with $\mu$ replaced by $\rho^{-1}$. The applied boundary condition is $\nabla p \cdot \normal=0$ on all boundaries. After multiplying by the test function $q \in P_1[\element]$, integrating over the domain, performing integration by parts, and including the SIP stabilisation terms, the result is
\begin{align}
  &- \int_\tess \nabla p \cdot\nabla q \dif\pos
  +\int_{\skel_I} q \nabla p \cdot\normal \dif s
  \\ \notag
  &+\int_{\skel_I} \jump{q} \avg{\nabla p }\cdot\normal^+ \dif s
  +\int_{\skel_I} \jump{\nabla p}\avg{ q}\cdot\normal^+ \dif s
  \\ \notag
  &+\int_{\skel_I} \jump{p} \avg{\nabla q}\cdot\normal^+ \dif s
  -\int_{\skel_I} \kappa_p \jump{p} \jump{q} \dif s = \int_\tess f q \dif\pos,
\end{align}
where $\skel_I$ is the set of all internal facets. The null space due to using pure Neumann boundary conditions is removed in the Krylov solver by providing the basis of the null space along with the system matrix \citep{PETSc_UM_2014}. The right-hand side will be the divergence of a known velocity field. This will be discretised as shown in \cref{eq:wf_ns_coupled2} to minimise the differences between the differential and the algebraic methods.

The main difference between solving for the pressure this way, by a forming an elliptic differential equation, instead of performing the elliptic-like algebraic operation used in the IPCS-A method described in \cref{sec:ipcs-d}, is the introduction of the penalty parameter. The need for stabilisation of elliptic operators in DG methods means that the resulting discrete elliptic matrices are different between the algebraic and the differential methods. The consequences of this are explored in \cref{sec:masscons} and shown numerically in \cref{sec:numexperiments}.

\section{The IPCS-D method}
\label{sec:ipcs-d}

The Incremental Pressure Correction Scheme on Differential form (IPCS-D) is an iterative version of the classic Chorin-Temam pressure correction scheme \citep{Chorin_1968,Temam_1969,ipcs_Goda_1979}. The method begins with solving the momentum equation for an approximate velocity field $\vel^*$ by use of a guessed pressure field, $p^*$,
\begin{align}
  \rho \left( \frac{1}{\Delta t}(\gamma_1\vel^* + \gamma_2\vel^n + \gamma_3\vel^{n-1}) + (\wvel\cdot\nabla) \vel^* \right) & = \nabla\cdot\mu\left(\nabla \vel^* + (\nabla\vel^*)^T\right) - \nabla p^*. \label{eq:IPCS_D_nsmomguess}
\end{align}

A splitting error is now introduced. Make the assumption that the true velocity $\vel$ and the velocity estimate $\vel^*$ are close enough so that
\begin{align}
  (\wvel\cdot\nabla)(\vel - \vel^*)                                            & \approx 0
  \label{eq:IPCS_D_spliterr1}
  \\
  \nabla\cdot\mu\left(\nabla(\vel - \vel^*) + (\nabla(\vel - \vel^*))^T\right) & \approx 0.
  \label{eq:IPCS_D_spliterr2}
\end{align}
By subtracting \cref{eq:IPCS_D_nsmomguess} from the true momentum \cref{eq:space_ns} and using the assumptions in \cref{eq:IPCS_D_spliterr1,eq:IPCS_D_spliterr2} the result is
\begin{align}
  \frac{\rho}{\Delta t}(\gamma_1\vel - \gamma_1\vel^*) = - \frac{1}{\rho}\nabla (p - p^*).
  \label{eq:IPCS_D_momred}
\end{align}

The velocity field should be divergence free, but the velocity estimate $\vel^*$ is not due to having used an approximate pressure field $p^*$, and not the true velocity field $p$. The continuity equation, $\nabla\cdot\vel=0$, is now used to eliminate the unknown velocity field $\vel$ from \cref{eq:IPCS_D_momred} by taking the divergence,
\begin{align}
  \nabla\cdot\left(\frac{\Delta t}{\gamma_1\rho}\nabla (p - p^*)\right) = \nabla\cdot \vel^*,
  \label{eq:IPCS_D_eqp}
\end{align}
which gives a Poisson equation for the unknown pressure $p$ with the guessed pressure $p^*$ and the velocity estimate $\vel^*$ as known coefficients. When the pressure $p$ is found it can be put back into \cref{eq:IPCS_D_momred} to find the updated velocity field, $\vel$.

\paragraph{The IPCS-D algorithm}
\begin{enumerate}
  \item Guess $p^*$ and solve \cref{eq:IPCS_D_nsmomguess} for $\vel^*$.
  \item Solve \cref{eq:IPCS_D_eqp} for $p$.
  \item Compute the updated velocity $\vel$ from \cref{eq:IPCS_D_momred}.
  \item Check for convergence by computing the residual $r_u = ||\V{u}^* - \V{u}||$ and optionally go to step 1 using $p$ as a new guess for the pressure if $r_u$ is not sufficiently small.
\end{enumerate}

\section{The IPCS-A method}
\label{sec:ipcs-a}

The Incremental Pressure Correction Scheme on Algebraic form (IPCS-A) starts from the Navier-Stokes equations on block matrix form, as can be seen in \cref{eq:saddle_point_problem}. The momentum equation can be solved for an approximate velocity field $\V{u}^*$ by use of a guessed pressure field $\V{p}^*$,
\begin{align}
  \M{A}\V{u}^* = \V{d} - \M{B}\V{p^*}.
  \label{eq:IPCSA_mom_with_guess}
\end{align}

A splitting error is now introduced. First let $\M{A}=\M{M}+\M{R}$ where $\M{M}$ is a scaled mass matrix resulting from the assembly of the first term in \cref{eq:op_A}, and $\M{R}$ contains the convective and diffusive operators. Then make the assumption that $\M{R}(\V{u} - \V{u}^*)\approx0$ and use this when subtracting \cref{eq:IPCSA_mom_with_guess} from the first line of \cref{eq:saddle_point_problem}, giving
\begin{align}
  \M{M}(\V{u} - \V{u}^*) = - \M{B}(\V{p} - \V{p}^*).
  \label{eq:IPCSA_spliterr}
\end{align}
The matrix $\M{M}$ is block diagonal and can easily be inverted. Use this property and the divergence free criterion, $\M{C}\V{u} = \V{e}$, to remove the unknown $\V{u}$ from \cref{eq:IPCSA_spliterr}
\begin{align}
  \V{e} - \M{C}\V{u}^* = - \M{C}\M{M}^{-1}\M{B}(\V{p} - \V{p}^*).
  \label{eq:IPCSA_soon_eqp}
\end{align}
Reorganising \cref{eq:IPCSA_soon_eqp} results in an equation for $\V{p}$
\begin{align}
  \M{C}\M{M}^{-1}\M{B}\V{p} = \M{C}\M{M}^{-1}\M{B}\V{p}^* - \V{e} + \M{C}\V{u}^*.
  \label{eq:IPCSA_eqp}
\end{align}

The velocity $\V{u}$ can be recovered without solving a linear system, simply by substituting the pressure $\V{p}$ from the solution of \cref{eq:IPCSA_eqp} into \cref{eq:IPCSA_spliterr},
\begin{align}
  \V{u} = - \M{M}^{-1}\M{B}(\V{p} - \V{p}^*).
  \label{eq:IPCSA_velup}
\end{align}

\paragraph{The IPCS-A algorithm}
\begin{enumerate}
  \item Guess $\V{p}^*$ and solve \cref{eq:IPCSA_mom_with_guess} for $\V{u}^*$.
  \item Solve \cref{eq:IPCSA_eqp} for $\V{p}$.
  \item Compute the updated velocity $\V{u}$ from \cref{eq:IPCSA_velup}.
  \item Check for convergence by computing the residual $r_u = ||\V{u}^* - \V{u}||$ and optionally go to step 1 using $\V{p}$ as a new guess for the pressure if $r_u$ is not sufficiently small.
\end{enumerate}

\section{The SIMPLE method}
\label{sec:simple}

This description is a summary of the method presented in \citet{KleinEtAl2013}, the Semi-Implicit Method for Pressure-Linked Equations, SIMPLE.
The method starts with the Navier-Stokes equations on algebraic block matrix form, \cref{eq:saddle_point_problem}. %
A guess $\V{p}^*$ is made and inserted into the governing equations which can be solved for an estimated velocity field, $\V{u}^*$,
\begin{align}
  \M{A} \V{u}^* & = \V{d} - \M{B} \V{p}^*,
  \label{eq:simple_momguess}
  \\
  \M{C} \V{u}^* & = \V{e} + \V{e}^*,
\end{align}
where $\V{e}^*$ is not zero when $\V{p}^*$ is not a perfect guess due to $\nabla \cdot \T{u}^*\neq 0$. Subtracting these equations from \cref{eq:saddle_point_problem} and defining corrections
$\Vh{u} = \V{u} - \V{u}^*$ and $\Vh{p} = \V{p} - \V{p}^*$ gives
\begin{align}
  \M{A} \Vh{u} & = -\M{B} \Vh{p},    \label{eq:SIMPLE_uhat}      \\
  \M{C} \Vh{u} & = -\V{e}^*.\label{eq:SIMPLE_phat}
\end{align}

A splitting error is now introduced. Construct a matrix $\tilde{\M{A}}  \approx \M{A}$ where $\tilde{\M{A}}$ approximates $\M{A}$, but is much easier to invert. A diagonal or block diagonal version of $\M{A}$ is used as an approximation in the numerical examples below. Note that this approximation allows the time derivative to be removed for steady-state problems, unlike in the IPCS-A method where we must require $\M{M}\neq0$. The velocity correction in \cref{eq:SIMPLE_uhat} can then be approximated as
\begin{align}
  \Vh{u} = - \tilde{\M{A}}^{-1} \M{B} \Vh{p},
  \label{eq:uupd}
\end{align}
and this approximation can be used to solve for $\Vh{p}$ by substitution into \cref{eq:SIMPLE_phat}
\begin{align}
  \left[\M{C} \tilde{\M{A}}^{-1} \M{B}\right]\Vh{p} = \V{e}^* = \M{C}\V{u}^* - \V{e}.
  \label{eq:solp}
\end{align}

The SIMPLE method is not guaranteed to converge without under-relaxation in the updates of $\V{p}$ and $\V{u}$ due to the approximate $\tilde{\M{A}}$. The under-relaxation of the pressure is performed explicitly,
\begin{align}
  \V{p} = \V{p}^* + \alpha_p \Vh{p},
  \label{eq:pupd}
\end{align}
while an implicit scheme is used for the velocity,
\begin{align}
  \left[\frac{1-\alpha_u}{\alpha_u} \tilde{\M{A}} + \M{A}\right] \V{u}^*
  =
  \V{d} - \M{B} \V{p}^*
  +
  \left[\frac{1-\alpha_u}{\alpha_u} \tilde{\M{A}}\right] \V{u}^*_\text{prev},
  \label{eq:solu}
\end{align}
with $0 < (\alpha_p,\alpha_u) < 1$.

\paragraph{The SIMPLE algorithm}
\begin{enumerate}
  \item Solve for $\V{u}^*$ using \cref{eq:solu} starting with guesses $\V{u}^*_\text{prev}$ and $\V{p}^*$
  \item Find $\Vh{p}$ using \cref{eq:solp}
  \item Update $\V{p}$ using \cref{eq:pupd}
  \item Update $\V{u}$ using \cref{eq:uupd}
  \item Check for convergence by computing the residual $r_u = ||\V{u}^* - \V{u}||$ and optionally go to step 1 with new
        guesses for $\V{u}^*$ and $\V{p}^*$
\end{enumerate}

\section{Exact mass conservation}
\label{sec:masscons}

The numerical representation of the velocity field will be divergence free to the precision of the discrete operators  if $\M{C}\V{u}-\V{e}=0$. In the IPCS-A method this is ensured on the algebraic level,
\begin{align}
  \M{C}\V{u} - \V{e}&=\  \M{C}\V{u} - \M{C}\V{u}^* + \M{C}\V{u}^* - \V{e} &
  \label{eq:masscons}
  \\ \notag
  & \overset{\mathclap{\text{\eqref{eq:IPCSA_spliterr}}}}=\ \M{C}\M{M}^{-1}\M{B}(\V{p} - \V{p}^*) + \M{C}\V{u}^* - \V{e}
  \\ \notag
  & \overset{\mathclap{\text{\eqref{eq:IPCSA_soon_eqp}}}}=\ 0.
\end{align}

The IPCS-D pressure Poisson \cref{eq:IPCS_D_eqp} is not equivalent to \cref{eq:IPCSA_soon_eqp} due to the stabilisation jump terms which will, even if the boundary conditions applied to \cref{eq:IPCS_D_eqp} are in perfect agreement with \cref{eq:IPCSA_eqp}, introduce a residual divergence in the velocity field since the assembled matrix from the elliptic operator with stabilisation does not satisfy \cref{eq:IPCSA_soon_eqp},
\begin{align}
  \text{Assemble}_{\text{\,DG},\kappa_p}\left(\nabla\cdot(\frac{\Delta t}{\gamma_1\rho}\nabla \parm)\right) \ne \M{C}\M{M}^{-1}\M{B}.
\end{align}
Other elliptic DG FEM discretisations like LDG and NIPG will also contain  stabilisation terms \citep{UnifiedDGElliptic2002}, and will hence have the same problem. A similar argument is shown in \citet{KleinEtAl2015}. They also show that the SIMPLE method is exactly divergence free with an argument similar to \cref{eq:masscons}.

\section{Numerical experiments}
\label{sec:numexperiments}

The weak form described in \cref{sec:discr} is implemented with quadratic discontinuous Galerkin elements for the velocity (DG2) and linear DG elements for the pressure (DG1). %
The numerical experiments are run with iterative solvers from PETSc 3.8 \citep{PETSc_UM_2014}. All results are from simulations running in parallel with MPI on the Abel computational cluster at the University of Oslo. The GMRES iterative solver has been used for both velocities and pressures and both the relative and absolute error criteria are \num{1e-15} with a maximum of \num{100} Krylov iterations. Convergence in the Krylov solver is typically achieved in less than \num{100} Krylov iterations for all but the first pressure correction iteration in the first time step.

An Additive Schwarz preconditioner (ASM) is used for the velocities and for the pressure an Algebraic Multigrid preconditioner (HYPRE BoomerAMG) is applied. The solvers are set up to use the previous solution as the initial guess. The Additive Schwarz preconditioner uses PETSc's default settings, which is incomplete LU factorization on each CPU. The BoomerAMG preconditioner also uses PETSc's default settings.

All the results shown below have been produced by Ocellaris \citep{landet_ocellaris_2019}, a single and multi-phase DG FEM Navier-Stokes solver. Input files for Ocellaris and post-processing scripts, which can be used to reproduce all figures shown in this paper, can be found in \citet{landet_zenodo_prscorr_2019}.

\subsection{Taylor-Green 2D flow}

The Taylor-Green vortex is an analytical solution to the 2D incompressible Navier-Stokes equations. The solution is
\begin{align}
  u & = -\sin(\pi y)\cos(\pi x)\exp(-2\pi^2\nu t), \notag                          \\
  v & = ~~\,\sin(\pi x) \cos(\pi y)\exp(-2\pi^2\nu t),                             \\
  p & = -\sfrac{1}{4}\,\rho(\cos{2\pi x} + \cos{2\pi y})\exp(-4\pi^2\nu t). \notag
\end{align}

\begin{figure}[htbp]
  \centering
  \includegraphics[width=0.70\textwidth]{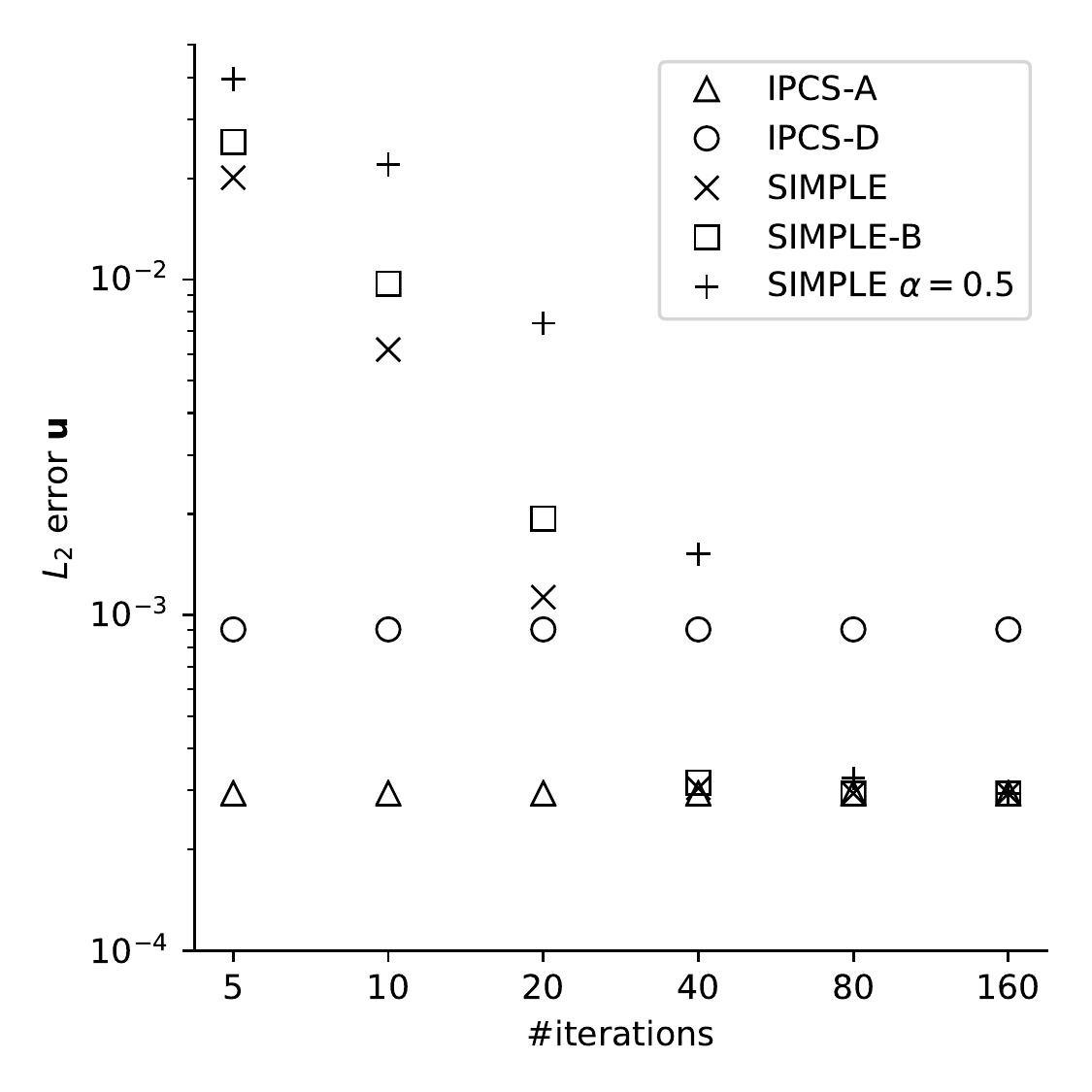}
  \caption{Taylor Green test case: effect of the number of pressure correction iterations per time step. The SIMPLE method is tested with a block-diagonal $\tilde{\M{A}}$ matrix (one block for each cell, `SIMPLE-B') in addition the the diagonal $\tilde{\M{A}}$ used for the rest of the simulations. A set of simulations with under-relaxation parameters $(\alpha_u, \alpha_p) = (0.5, 0.5)$ is included in addition to the $(\alpha_u, \alpha_p) = (0.7, 1.0)$ choice used for the rest of the simulations. }
  \label{fig:tg_niters}
\end{figure}

Spatial convergence of the pressure correction methods is examined on a regular grid. The domain, $(x, y) \in [0, 2] \times [0, 2]$, is divided into $N \times N$ rectangles that are then subdivided into two triangles each. Grid sizes $N=\{8, 16, 24, 32\}$ are used to study the spatial convergence rate. The time step is $\Delta t= 0.01$ and the numerical results are compared to the analytical expressions at $t=1.0$. Exactly 160 pressure correction iterations are performed for each time step, which is sufficient for all methods, see \cref{fig:tg_niters}. The physical parameters applied are $\rho=1.0$ and $\nu=0.005$.

In the SIMPLE method an under-relaxation factor of $\alpha_u=0.7$ is used for the velocity and $\alpha_p=1.0$ is used for the pressure. This is in line with \citet{KleinEtAl2013}. Setting both under-relaxation factors to \num{1.0} (no under-relaxation) as was done in \citet{KleinEtAl2015} is unstable for the current simulations and causes the iteration procedure to blow up. A diagonal $\tilde{\M{A}}$ matrix is used with no lumping.

The number of pressure correction iterations required to reach convergence is shown in \cref{fig:tg_niters}. The SIMPLE method requires far more iterations in order to converge than the IPCS methods. The influence of some of the tunable parameters in the SIMPLE method are also shown in the figure. Some sensitivity is found both in regard to how the approximate inverse of $\M{A}$ is computed and in the choice of under-relaxation parameters. The effect of these choices are significantly smaller than the effect of changing the pressure correction method.

\begin{figure}[htbp]
  \centering
  \subcaptionbox{Velocity\label{fig:tg_vel}}{\includegraphics[width=0.4\textwidth]{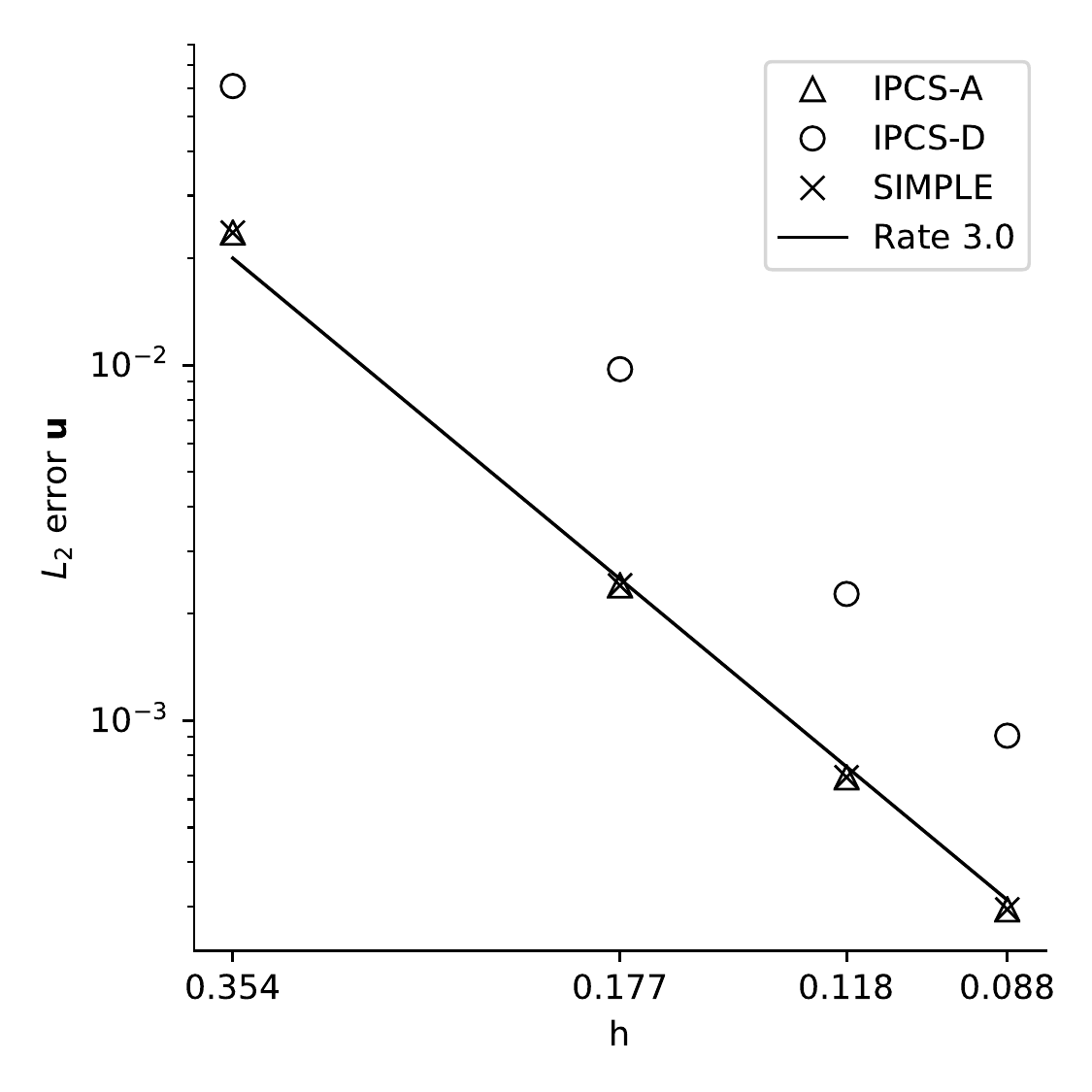}}
  \hspace{0.1\textwidth}
  \subcaptionbox{Pressure\label{fig:tg_prs}}{\includegraphics[width=0.4\textwidth]{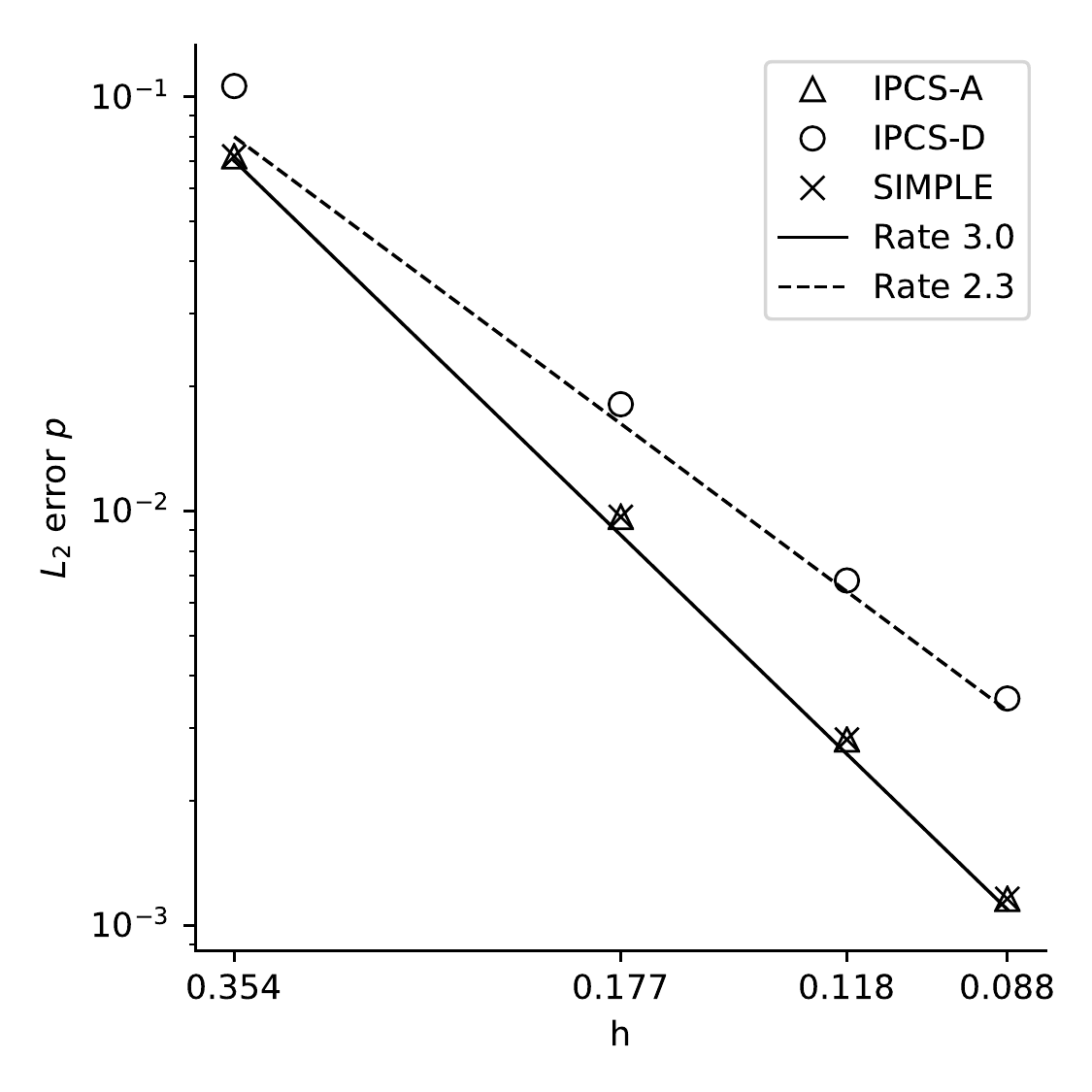}}
  \\
  \subcaptionbox{Divergence of $\vel$ in DG0\label{fig:tg_div}}{\includegraphics[width=0.4\textwidth]{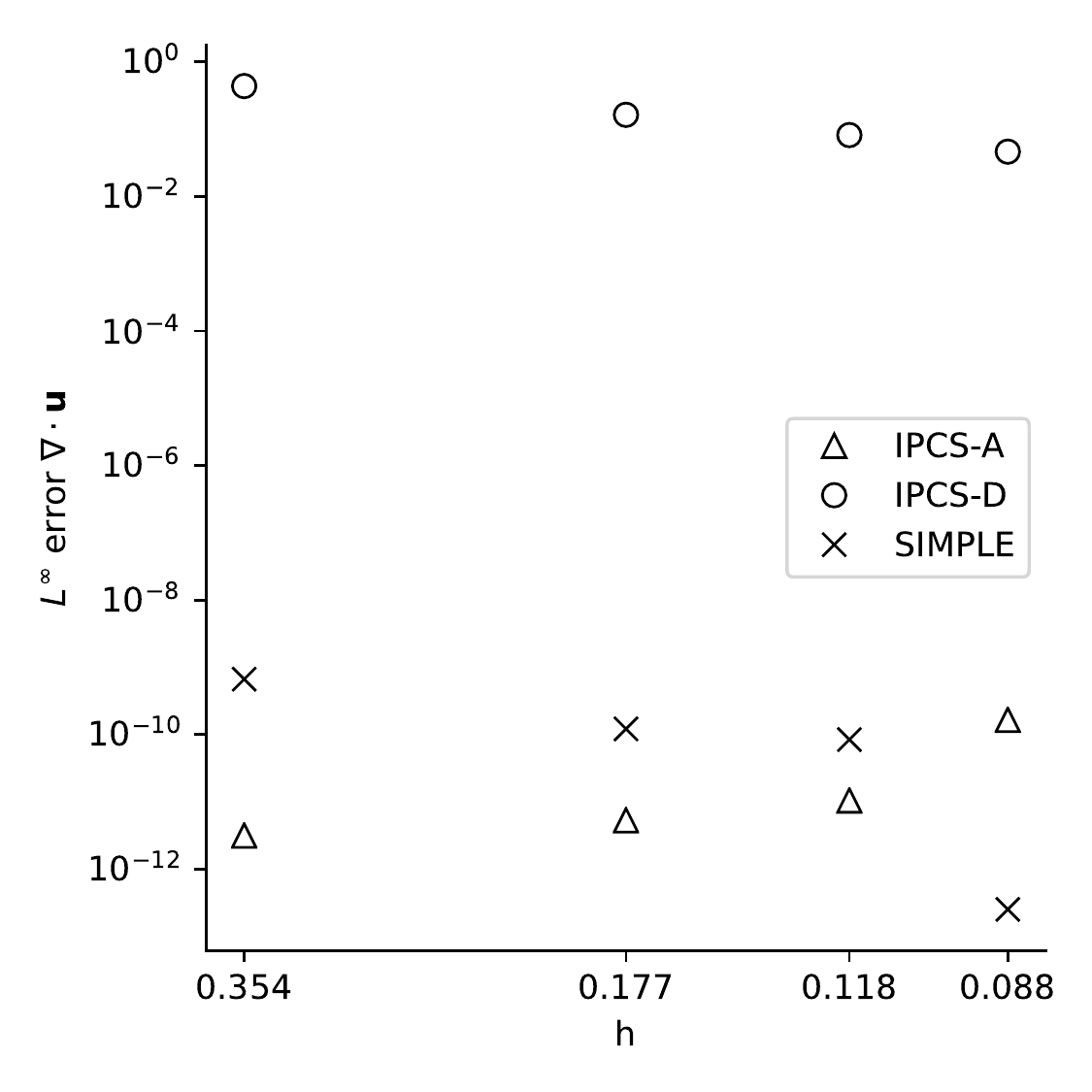}}
  \hspace{0.1\textwidth}
  \subcaptionbox{Divergence of $\vel$ in DG2\label{fig:tg_div2}}{\includegraphics[width=0.4\textwidth]{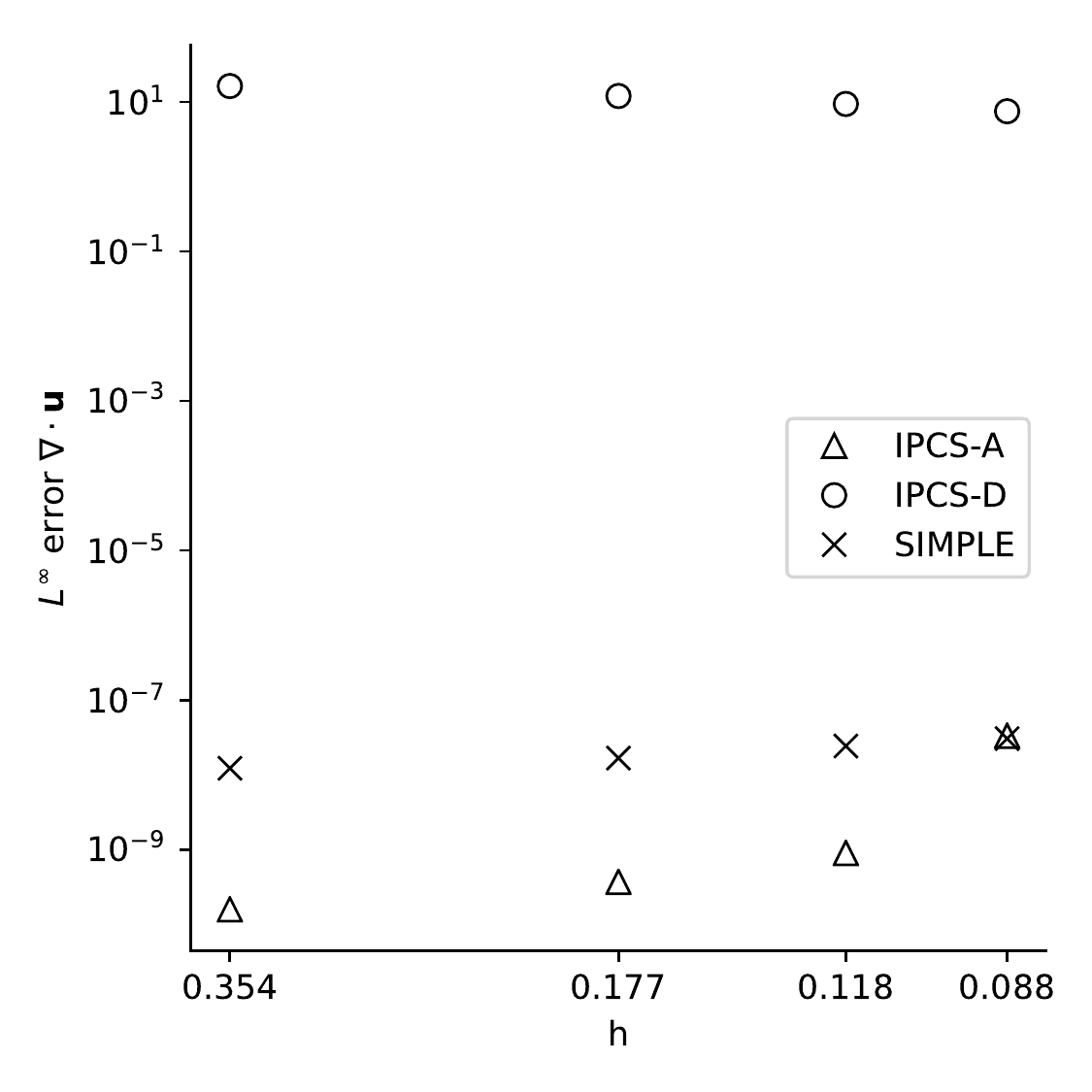}}
  \caption{Taylor Green test case: spatial convergence.}
  \label{fig:tg_space}
\end{figure}

Spatial convergence rates of the velocity and pressure can be seen in \cref{fig:tg_vel,fig:tg_prs}. The velocity converges with the expected order while the pressure super-converges due to the regular mesh. It is normal to observe super-convergence on highly regular meshes \citep{superconvergence_2012}. The divergence of the velocity field is shown in \cref{fig:tg_div}. The divergence has been projected to a piecewise constant (DG0) function space before computing the $L_2$ norm. The IPCS-D method gives a velocity field with approximately ten orders of magnitude higher divergence than IPCS-A and SIMPLE. The divergence is also computed in the space of the velocity, piecewise quadratics (DG2), and this is shown in \cref{fig:tg_div2}. The same behaviour is shown here, though the error is higher.

\begin{figure}[htbp]
  \centering
  \subcaptionbox{Velocity\label{fig:tg_time_vel}}{\includegraphics[width=0.4\textwidth]{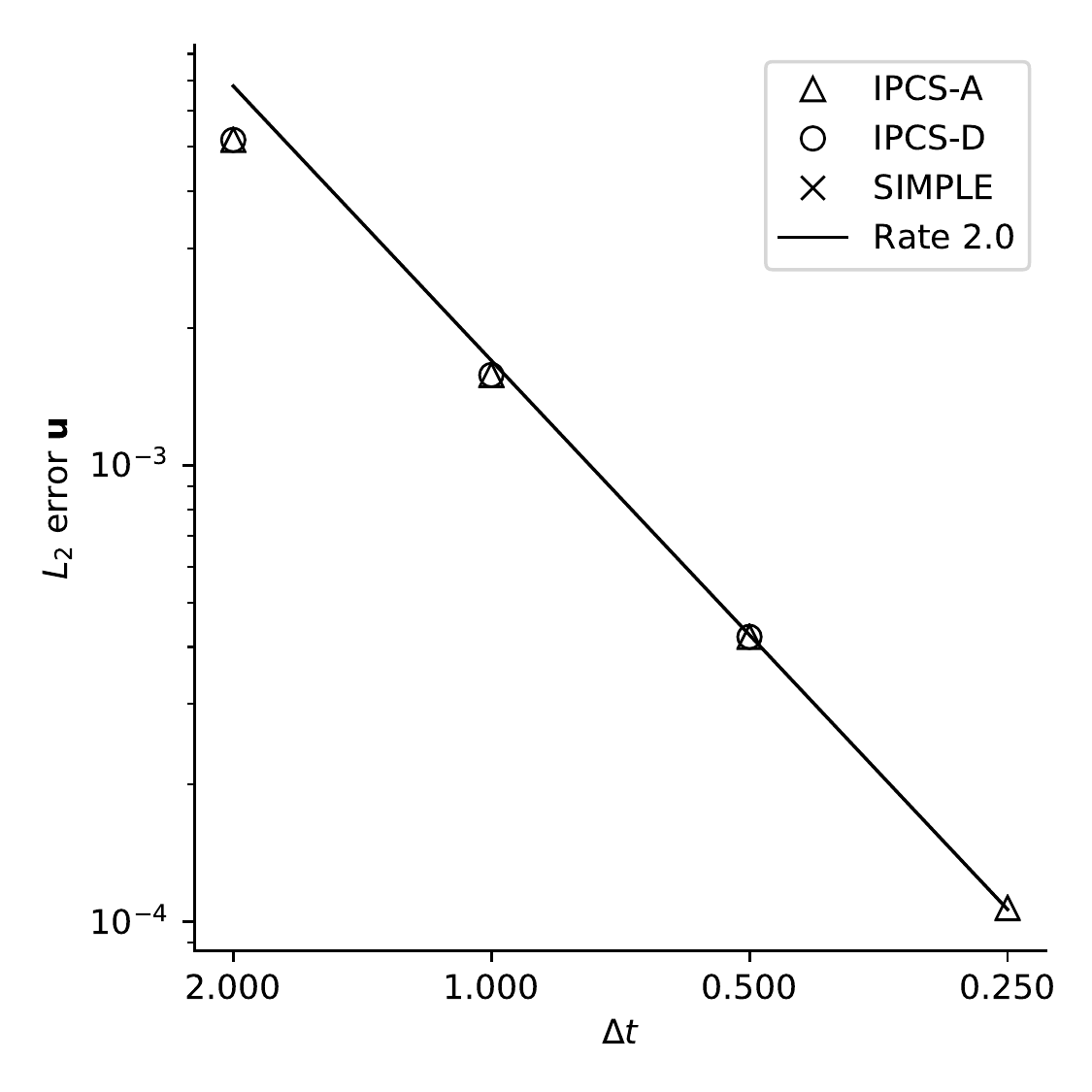}}
  \hspace{0.1\textwidth}
  \subcaptionbox{Pressure\label{fig:tg_time_prs}}{\includegraphics[width=0.4\textwidth]{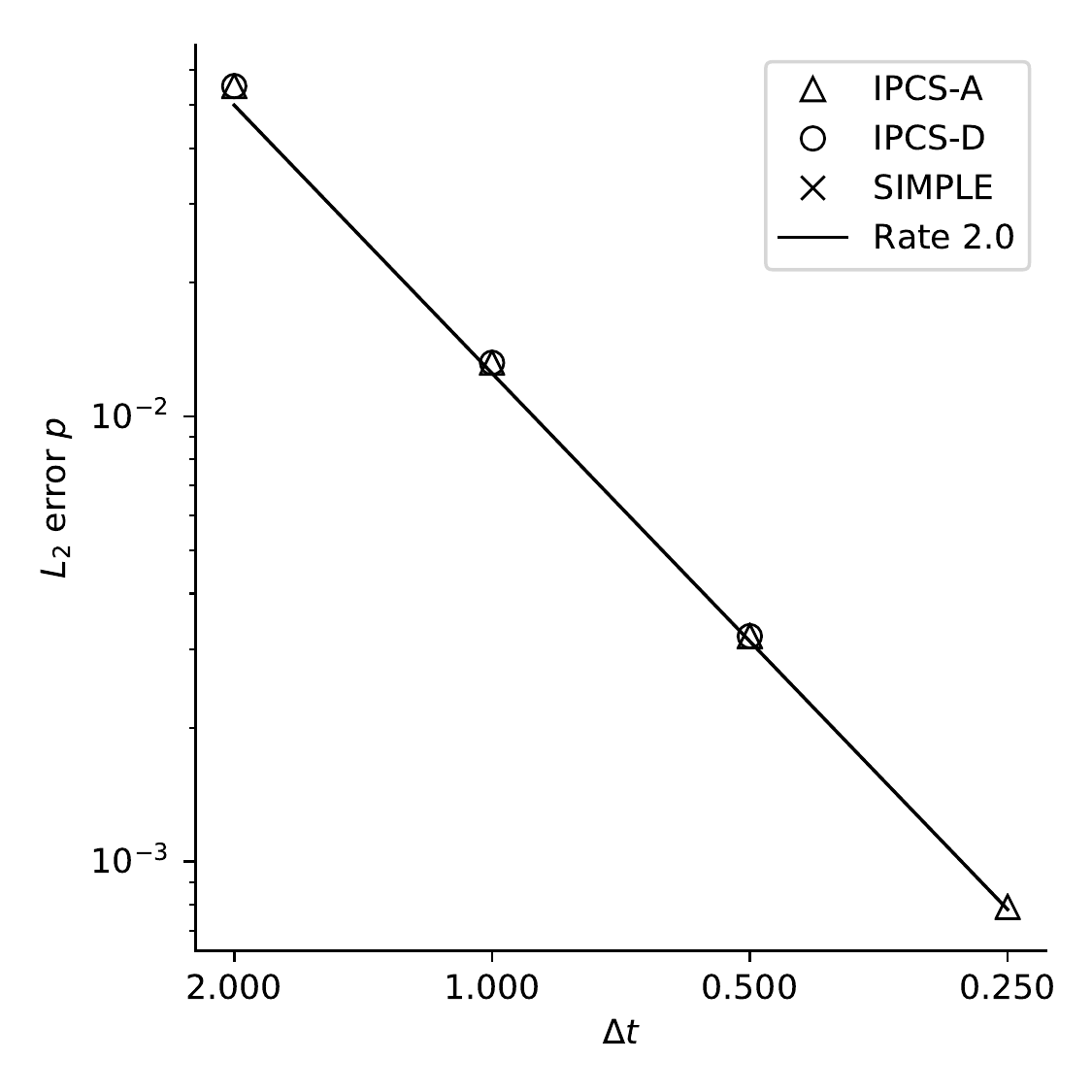}}
  \caption{Taylor Green test case: temporal convergence.}
  \label{fig:tg_time}
\end{figure}

Studying temporal convergence for the selected Taylor-Green flow requires a very fine spatial discretisation since the error in the time stepping routine is much smaller than the spatial error. The results obtained with a very fine grid where $N=200$ is shown in \cref{fig:tg_time}. The time step is gradually refined with $\Delta t = \{2.00, 1.00, 0.50, 0.25\}$ and the numerical  solution is compared with the analytical at $t=6.0$. The number of pressure correction iterations per time step is \num{200} and the number of Krylov solver iterations is also set to \num{200}. This is a necessary increase to achieve the theoretical convergence rate for the velocity. The pressure will converge at a rate slightly above \num{2.0} with fewer iterations, but even with the fine spatial discretisation, the convergence rate of the velocity is slightly below \num{2.0} and the Krylov solver is never fully converged even after \num{200} GMRES iterations per pressure iteration. Between refinements $\Delta t=0.5$ and $\Delta t=0.25$ the observed convergence rate of the velocity is \num{1.97}.

\subsection{Ethier-Steinman 3D flow}

An analytical solution to the 3D incompressible Navier-Stokes equations is given by \citet[equation 15]{Ethier_Steinman_1994} as
\begin{align}
  u & = -a e^{-b^2\nu t}[e^{ax}\sin(ay + bz) + e^{az}\cos(ax + by)], \notag \\
  v & = -a e^{-b^2\nu t}[e^{ay}\sin(az + bx) + e^{ax}\cos(ay + bz)], \notag \\
  v & = -a e^{-b^2\nu t}[e^{az}\sin(ax + by) + e^{ay}\cos(az + bx)], \notag \\
  p & = \frac{-a^2}{2} e^{-2b^2\nu t} [
  e^{2ax} + e^{2ay} + e^{2az} - \bar{p}
  \label{eq:et_analytical}
  \\ &\qquad\qquad\qquad+
  2\sin(ax + by)\cos(az + bx)e^{a(y + z)}
  \notag                                                                     \\ &\qquad\qquad\qquad+
  2\sin(ay + bz)\cos(ax + by)e^{a(z + x)}
  \notag                                                                     \\ &\qquad\qquad\qquad+
  2\sin(az + bx)\cos(ay + bz)e^{a(x + y)}]. \notag
\end{align}

A constant correction for the pressure, $\bar{p}$, is included in \cref{eq:et_analytical} to ensure that the average pressure is zero. This is enforced in the linear solver to remove the pressure kernel from the system. On a domain $(x, y, z) \in [-1,1]\times[-1,1]\times[-1,1]$ this gives
\begin{align}
  \bar{p} =  [-15\pi^2 - 32 + 32e^\pi + 15\pi^2e^\pi](e^\frac{\pi}{2} 5\pi^3)^{-1}.
\end{align}

Dirichlet boundary conditions are applied to the velocity.
The domain is divided into a regular mesh with $N \times N \times N$ cubes which are further subdivided into six tetrahedra in each cube. The time step is $\Delta t = 0.001$ and the numerical solution is compared to the analytical at time $t=0.1$.

\begin{figure}[htbp]
  \centering
  \subcaptionbox{Velocity\label{fig:es_vel}}{\includegraphics[width=0.4\textwidth]{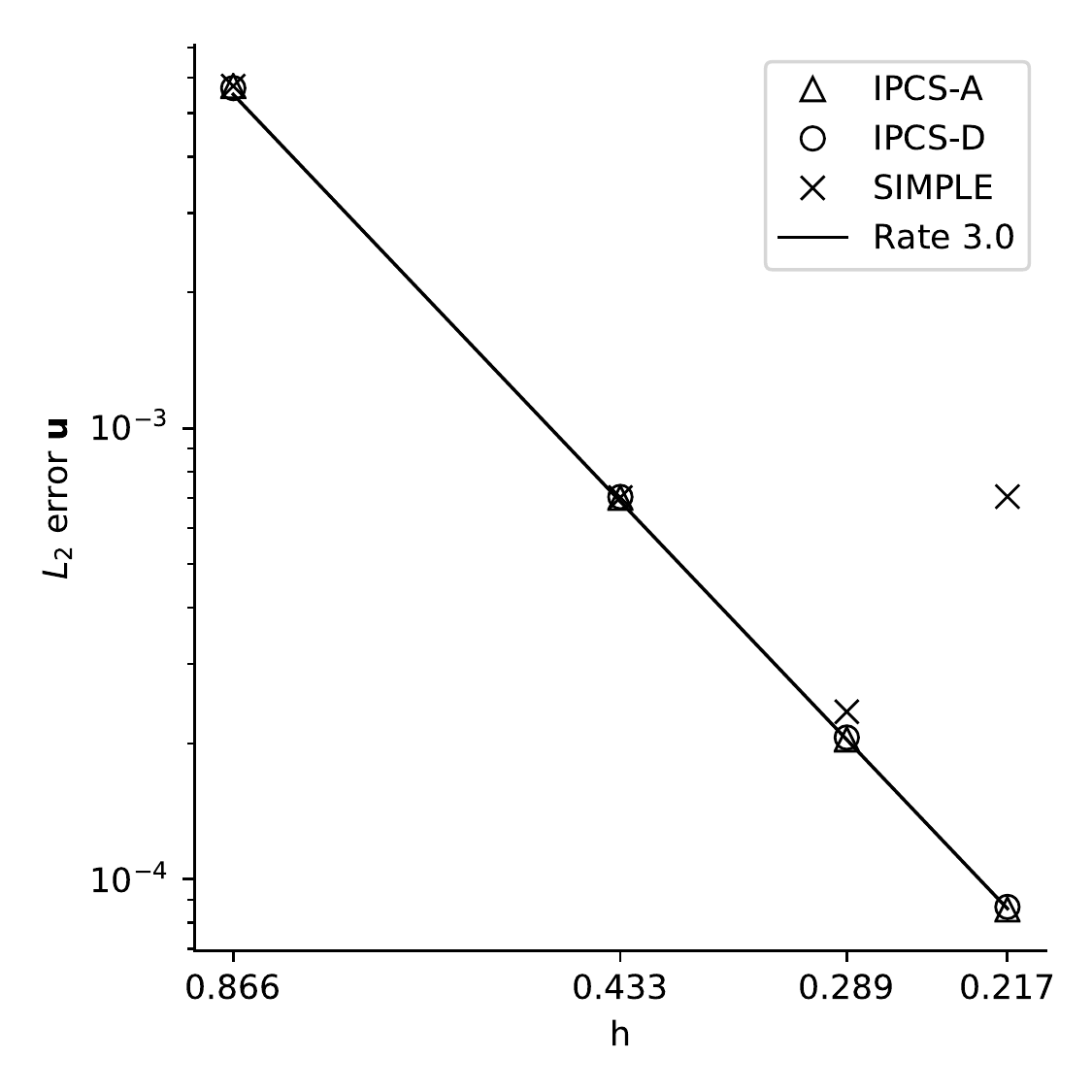}}
  \hspace{0.1\textwidth}
  \subcaptionbox{Pressure\label{fig:es_prs}}{\includegraphics[width=0.4\textwidth]{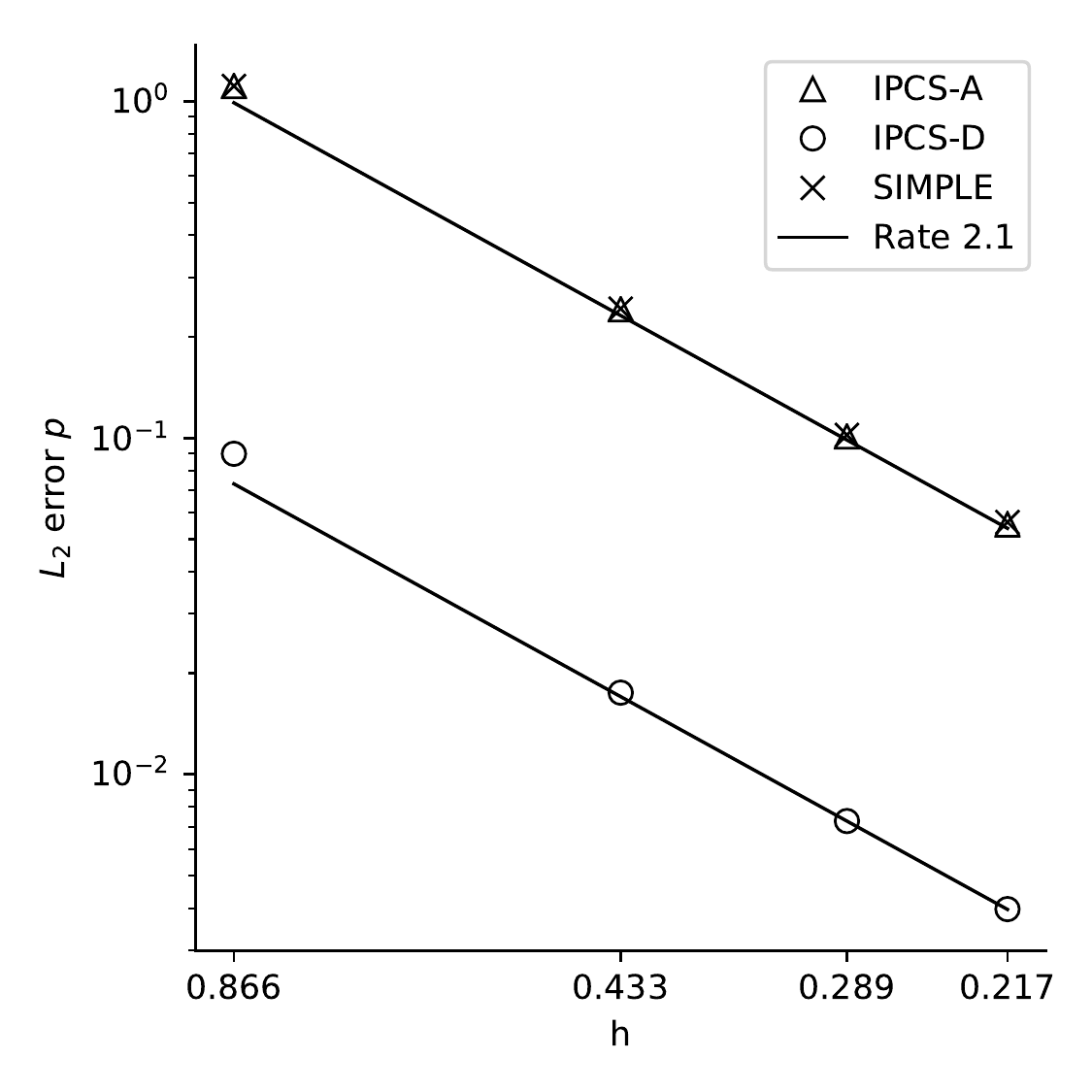}}
  \subcaptionbox{Divergence of $\vel$ in DG0 \label{fig:es_div0}}{\includegraphics[width=0.4\textwidth]{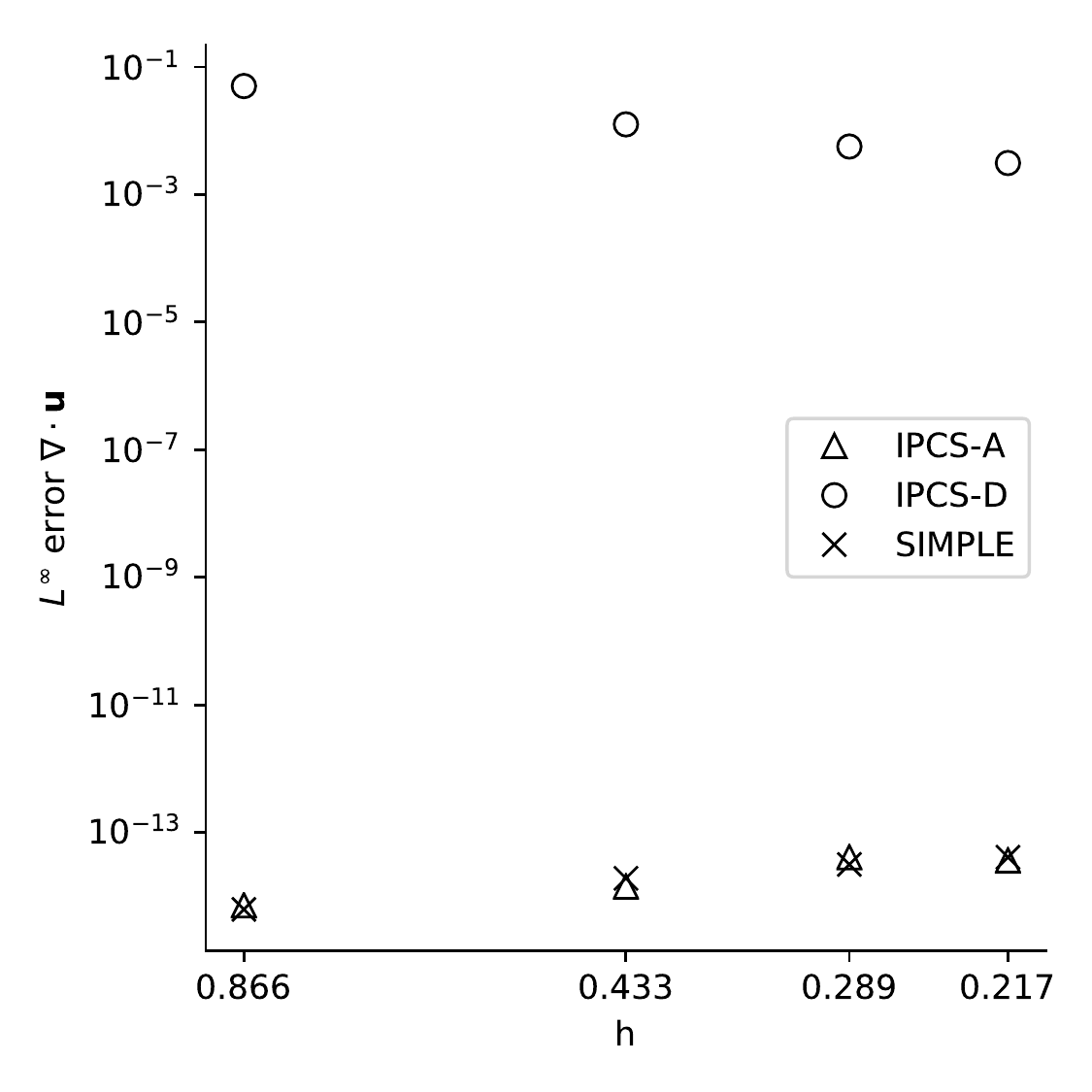}}
  \caption{Ethier-Steinman test case: spatial convergence.}
  \label{fig:es_space}
\end{figure}

The spatial convergence of the methods in Ethier-Steinman flow can be seen in \cref{fig:es_space}. The results for the IPCS methods are presented with 100 pressure correction iterations per time step, but the results differ only in the third significant digit from the results obtained with only 5 iterations. The SIMPLE method has been run with $\alpha_u=\alpha_p=0.5$ to get stable iterations at the finer grid resolutions where using $\alpha_u=0.7$ and $\alpha_p=1.0$ would blow up. The SIMPLE results are produced using 500 pressure correction iterations per time step. This gives an optimal convergence rate for the pressure at all resolutions and for the velocity at all but the finest resolution. With 100 pressure correction iterations only two points would fall on the line in \cref{fig:es_vel}. The divergence error norms for the algebraic methods are again close to machine precision. The IPCS-D divergence error is much higher, but getting smaller with increasing mesh resolution, as expected.

\subsection{Efficiency}

\begin{figure}[htbp]
  \centering
  \includegraphics[width=0.55\textwidth]{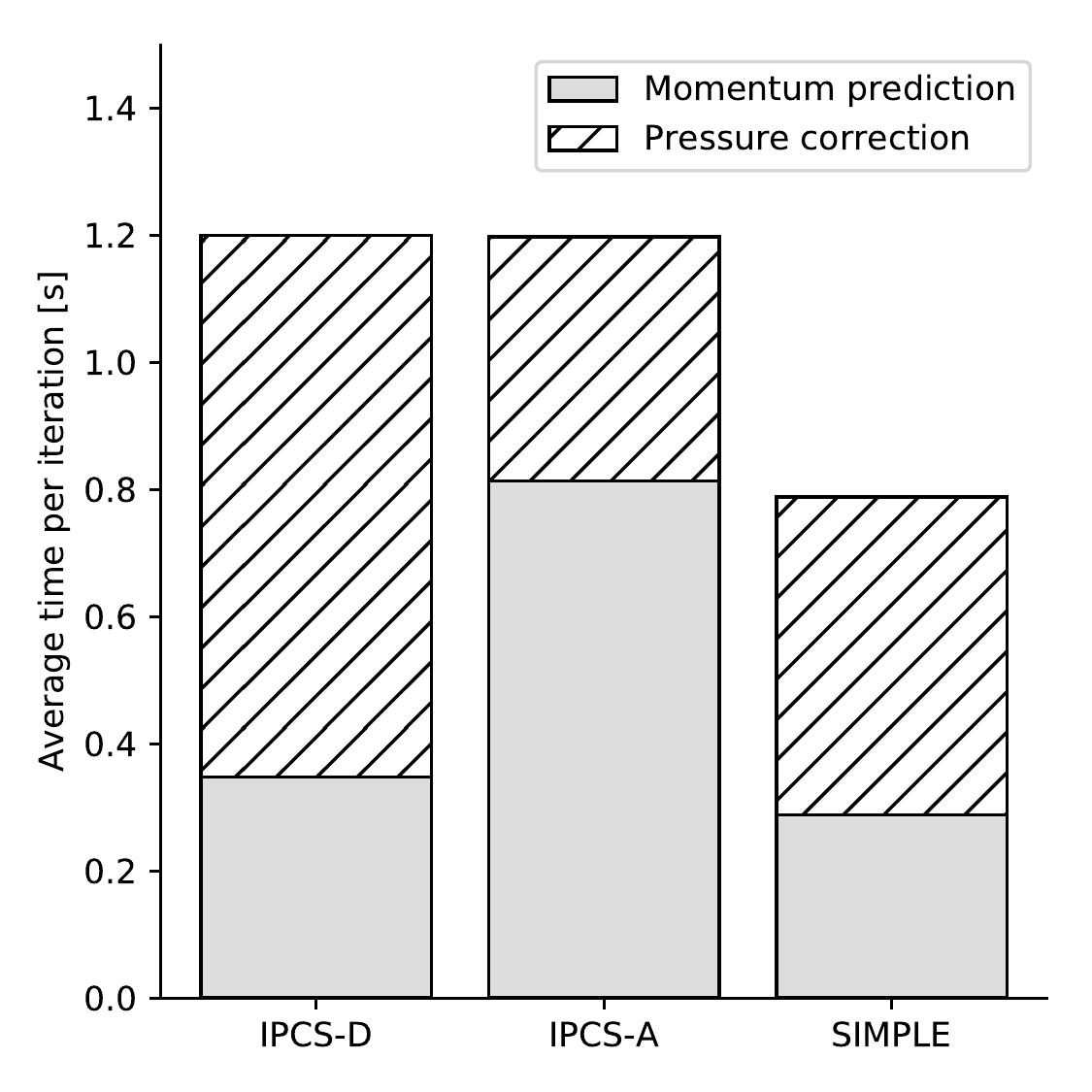}
  \caption{The average wall clock time spent in the two pressure correction steps. Ethier-Steinman, $N$=12, 100 pressure corrections in each of the 100 time steps, 16 CPUs. The minimum average times from five full simulations of each method have been used. The noticeable differences between the three methods are very repeatable.}
  \label{fig:es_wallclock}
\end{figure}

Producing a fair wall clock running time comparison of the pressure correction methods is more tricky than reporting convergence results. For the convergence rates it only matters that the Krylov iterations inside each pressure correction iteration converge in the end, not how many iterations are needed for convergence. The total running time is totally dominated by the time spent in the momentum prediction and pressure correction steps, and except for matrix assembly, which takes around \num{1}{\%} of the wall clock time, this is all time spent in the Krylov solvers. Tweaking the preconditioner parameters used for the Krylov solvers in each method may improve one of the methods more than the other, causing this comparison---which uses the same preconditioner parameters for all three methods---to not be very relevant. With that caveat in mind, the time spent in the two steps can be seen in \cref{fig:es_wallclock}.

The total time per pressure correction iteration is less in the SIMPLE method than in the IPCS methods, which is to be expected due to the under-relaxation procedure causing the unknown vectors to be closer to the initial guesses. The IPCS-D method spends more time in the pressure correction than the IPCS-A method. %
The reason may be that the penalty parameter, which ensures coercivity, also makes the resulting matrix more and more ill-conditioned as the penalty parameter increases, although the parameter we have used, \cref{eq:penalty_param}, should be close to optimal. The momentum prediction step takes significantly more time in IPCS-A than in IPCS-D. The reason may be related to the skew symmetric convective term, \cref{eq:wf_ns_skewsymm}, making the IPCS-D $\M{A}$ matrix more diagonal dominant. Unfavourable right hand sides may also require more iterations.

\section{Conclusions}

For both the 2D (Taylor-Green, \cref{fig:tg_space}) and 3D (Ethier-Steinman, \cref{fig:es_space}) numerical experiments, the convergence results show that the algebraic pressure splitting methods are exactly divergence free while the differential IPCS-D method is not. This is in line with the expected results from the discussion in \cref{sec:masscons}. For both 2D and 3D tests the IPCS methods converge perfectly with 5 pressure correction iterations per time step, while the SIMPLE method requires between 10 and 100 times more iterations to produce converged results. The SIMPLE method spends slightly less time per iteration (\cref{fig:es_wallclock}), but is not fast enough to outweigh the vastly increased number of iterations required to obtain converged results. For mass conserving pressure correction iterations the IPCS-A method is most efficient, and hence the recommended option among the three tested methods.

\section*{Acknowledgements}

The simulations were performed on resources provided by UNINETT Sigma2, the National Infrastructure for High Performance Computing and Data Storage in Norway.

\bibliographystyle{plainnat}
\bibliography{references}

\begin{thebibliography}{34}
\providecommand{\natexlab}[1]{#1}
\providecommand{\url}[1]{\texttt{#1}}
\expandafter\ifx\csname urlstyle\endcsname\relax
  \providecommand{\doi}[1]{doi: #1}\else
  \providecommand{\doi}{doi: \begingroup \urlstyle{rm}\Url}\fi

\bibitem[Amestoy et~al.(2001)Amestoy, Duff, Koster, and L'Excellent]{MUMPS:1}
P.~R. Amestoy, I.~S. Duff, J.~Koster, and J.-Y. L'Excellent.
\newblock A fully asynchronous multifrontal solver using distributed dynamic
  scheduling.
\newblock \emph{SIAM Journal on Matrix Analysis and Applications}, 23\penalty0
  (1):\penalty0 15--41, 2001.

\bibitem[Amestoy et~al.(2006)Amestoy, Guermouche, L'Excellent, and
  Pralet]{MUMPS:2}
P.~R. Amestoy, A.~Guermouche, J.-Y. L'Excellent, and S.~Pralet.
\newblock Hybrid scheduling for the parallel solution of linear systems.
\newblock \emph{Parallel Computing}, 32\penalty0 (2):\penalty0 136--156, 2006.

\bibitem[Arnold(1982)]{arnold_interior_1982}
Douglas~Norman Arnold.
\newblock An interior penalty finite element method with discontinuous
  elements.
\newblock \emph{SIAM journal on numerical analysis}, 19\penalty0 (4):\penalty0
  742--760, 1982.

\bibitem[Arnold et~al.(2002)Arnold, Brezzi, Cockburn, and
  Marini]{UnifiedDGElliptic2002}
Douglas~Norman Arnold, Franco Brezzi, Bernardo Cockburn, and Donatella Marini.
\newblock Unified analysis of discontinuous {Galerkin} methods for elliptic
  problems.
\newblock \emph{SIAM J. Numer. Anal.}, 39\penalty0 (5):\penalty0 1749–1779,
  2002.

\bibitem[Balay et~al.(2014)Balay, Abhyankar, Adams, Brown, Brune, Buschelman,
  Eijkhout, Gropp, Kaushik, Knepley, and et~al.]{PETSc_UM_2014}
Satish Balay, Shrirang Abhyankar, Mark~F. Adams, Jed Brown, Peter Brune, Kris
  Buschelman, Victor Eijkhout, William~D. Gropp, Dinesh Kaushik, Matthew~G.
  Knepley, and et~al.
\newblock \emph{{PETSc} Users Manual}.
\newblock Number ANL-95/11-Revision 3.5. 2014.

\bibitem[Chorin(1968)]{Chorin_1968}
Alexandre~Joel Chorin.
\newblock Numerical solution of the {Navier-Stokes} equations.
\newblock \emph{Mathematics of computation}, 22\penalty0 (104):\penalty0
  745–762, 1968.

\bibitem[Cockburn et~al.(2005)Cockburn, Kanschat, and
  Schötzau]{cockburn_locally_2005}
Bernardo Cockburn, Guido Kanschat, and Dominik Schötzau.
\newblock A locally conservative {LDG} method for the incompressible
  {Navier}-{Stokes} equations.
\newblock \emph{Mathematics of Computation}, 74\penalty0 (251):\penalty0
  1067--1096, 2005.

\bibitem[Epshteyn and Rivière(2007)]{epshteyn_estimation_2007}
Yekaterina Epshteyn and Béatrice Rivière.
\newblock Estimation of penalty parameters for symmetric interior penalty
  {Galerkin} methods.
\newblock \emph{Journal of Computational and Applied Mathematics}, 206\penalty0
  (2):\penalty0 843--872, 2007.

\bibitem[Ethier and Steinman(1994)]{Ethier_Steinman_1994}
C.~Ross Ethier and D.~A. Steinman.
\newblock Exact fully {3D} {Navier–Stokes} solutions for benchmarking.
\newblock \emph{International Journal for Numerical Methods in Fluids}, 1994.

\bibitem[Fletcher(1976)]{bi-cg76}
R.~Fletcher.
\newblock Conjugate gradient methods for indefinite systems.
\newblock In G.~Alistair Watson, editor, \emph{Numerical Analysis}, Lecture
  Notes in Mathematics, pages 73--89. Springer Berlin Heidelberg, 1976.

\bibitem[Goda(1979)]{ipcs_Goda_1979}
Katuhiko Goda.
\newblock A multistep technique with implicit difference schemes for
  calculating two- or three-dimensional cavity flows.
\newblock \emph{Journal of Computational Physics}, 30\penalty0 (1):\penalty0
  76--95, 1979.

\bibitem[Gresho(1991)]{gresho_cfd_issues_1991}
P.~M. Gresho.
\newblock Some current {CFD} issues relevant to the incompressible
  navier-stokes equations.
\newblock \emph{Computer Methods in Applied Mechanics and Engineering},
  87\penalty0 (2):\penalty0 201--252, 1991.

\bibitem[Guermond and Shen(2003)]{guermond_velocity-correction_2003}
J.~Guermond and J.~Shen.
\newblock Velocity-correction projection methods for incompressible flows.
\newblock \emph{{SIAM} Journal on Numerical Analysis}, 41\penalty0
  (1):\penalty0 112--134, 2003.

\bibitem[Guillén-gonzález and Tierra(2012)]{superconvergence_2012}
Francisco Guillén-gonzález and Giordano Tierra.
\newblock Superconvergence in velocity and pressure for the 3d time-dependent
  {Navier}-{Stokes} equations.
\newblock \emph{SeMA Journal}, 57\penalty0 (1):\penalty0 49--67, 01 2012.

\bibitem[Hestenes and Stiefel(1952)]{cgmethod52}
Magnus~Rudolph Hestenes and Eduard Stiefel.
\newblock \emph{Methods of conjugate gradients for solving linear systems},
  volume~49.
\newblock {NBS} Washington, {DC}, 1952.

\bibitem[Issa(1986)]{Issa1986}
Raad Issa.
\newblock Solution of the implicitly discretised fluid flow equations by
  operator-splitting.
\newblock \emph{Journal of Computational Physics}, 62\penalty0 (1):\penalty0
  40–65, 1986.

\bibitem[Kawahara and Ohmiya(1985)]{kawahara_velcor_1985}
Mutsuto Kawahara and Kiyotaka Ohmiya.
\newblock Finite element analysis of density flow using the velocity correction
  method.
\newblock \emph{International Journal for Numerical Methods in Fluids},
  5\penalty0 (11):\penalty0 981--993, 1985.

\bibitem[Klein et~al.(2013)Klein, Kummer, and Oberlack]{KleinEtAl2013}
Benedikt Klein, Florian Kummer, and Martin Oberlack.
\newblock A {SIMPLE} based discontinuous {Galerkin} solver for steady
  incompressible flows.
\newblock \emph{Journal of Computational Physics}, 237:\penalty0 235–250, 03
  2013.

\bibitem[Klein et~al.(2015)Klein, Kummer, Keil, and Oberlack]{KleinEtAl2015}
Benedikt Klein, Florian Kummer, Markus Keil, and Martin Oberlack.
\newblock An extension of the {SIMPLE} based discontinuous {Galerkin} solver to
  unsteady incompressible flows.
\newblock \emph{International Journal for Numerical Methods in Fluids},
  77\penalty0 (10):\penalty0 571–589, 04 2015.

\bibitem[Klein et~al.(2016)Klein, Müller, Kummer, and Oberlack]{KleinEtAl2016}
Benedikt Klein, Björn Müller, Florian Kummer, and Martin Oberlack.
\newblock A high-order discontinuous {Galerkin} solver for low {Mach} number
  flows.
\newblock \emph{International Journal for Numerical Methods in Fluids},
  81\penalty0 (8):\penalty0 489–520, 07 2016.

\bibitem[Landet(2019{\natexlab{a}})]{landet_ocellaris_2019}
Tormod Landet.
\newblock The ocellaris finite element solver for free surface flows,
  2019{\natexlab{a}}.
\newblock URL \url{https://www.ocellaris.org/}.
\newblock \href{https://www.ocellaris.org/}{www.ocellaris.org}.

\bibitem[Landet(2019{\natexlab{b}})]{landet_zenodo_prscorr_2019}
Tormod Landet.
\newblock Input files and plots, 2019{\natexlab{b}}.
\newblock URL \url{http://dx.doi.org/10.5281/zenodo.2556909}.
\newblock \href{http://doi.org/10.5281/zenodo.2556909}{Zenodo:
  10.5281/zenodo.2556909}.

\bibitem[Landet et~al.(2018)Landet, Mardal, and Mortensen]{landet_slope_2018}
Tormod Landet, Kent-Andre Mardal, and Mikael Mortensen.
\newblock Slope limiting the velocity field in a discontinuous {Galerkin}
  divergence free two-phase flow solver.
\newblock \emph{{arXiv}:\href{http://arxiv.org/abs/1803.06976}{1803.06976
  [physics]}}, 2018.
\newblock URL \url{http://arxiv.org/abs/1803.06976}.

\bibitem[Li(2005)]{superlu05}
Xiaoye~S. Li.
\newblock An overview of {SuperLU}: Algorithms, implementation, and user
  interface.
\newblock \emph{ACM Transactions on Mathematical Software}, 31\penalty0
  (3):\penalty0 302--325, 09 2005.

\bibitem[Li and Demmel(2003)]{superlu_dist03}
Xiaoye~S. Li and James~W. Demmel.
\newblock {SuperLU\_DIST}: A scalable distributed-memory sparse direct solver
  for unsymmetric linear systems.
\newblock \emph{{ACM} Trans. Math. Softw.}, 29\penalty0 (2):\penalty0 110--140,
  2003.

\bibitem[Li et~al.(1999)Li, Demmel, Gilbert, iL. Grigori, Shao, and
  Yamazaki]{superlu_ug99}
X.S. Li, J.W. Demmel, J.R. Gilbert, iL. Grigori, M.~Shao, and I.~Yamazaki.
\newblock {SuperLU Users' Guide}.
\newblock Technical Report LBNL-44289, Lawrence Berkeley National Laboratory,
  09 1999.
\newblock
  \href{http://crd.lbl.gov/~xiaoye/SuperLU/}{crd.lbl.gov/\textasciitilde{}xiaoye/SuperLU}.

\bibitem[Patankar and Spalding(1972)]{simple_1972}
S.~V Patankar and D.~B Spalding.
\newblock A calculation procedure for heat, mass and momentum transfer in
  three-dimensional parabolic flows.
\newblock \emph{International Journal of Heat and Mass Transfer}, 15\penalty0
  (10):\penalty0 1787--1806, 1972.

\bibitem[Saad and Schultz(1986)]{gmres86}
Y.~Saad and M.~Schultz.
\newblock {GMRES}: A generalized minimal residual algorithm for solving
  nonsymmetric linear systems.
\newblock \emph{{SIAM} Journal on Scientific and Statistical Computing},
  7\penalty0 (3):\penalty0 856--869, 1986.

\bibitem[Schur(1917)]{schur_1917}
Issai Schur.
\newblock Über potenzreihen, die im innern des einheitskreises beschränkt
  sind.
\newblock \emph{Journal für die reine und angewandte Mathematik},
  147:\penalty0 205--232, 1917.

\bibitem[Shahbazi et~al.(2007)Shahbazi, Fischer, and Ethier]{shahbazi_2007}
Khosro Shahbazi, Paul~F. Fischer, and C.~Ross Ethier.
\newblock A high-order discontinuous galerkin method for the unsteady
  incompressible navier–stokes equations.
\newblock \emph{Journal of Computational Physics}, 222\penalty0 (1):\penalty0
  391--407, 2007.

\bibitem[Temam(1969)]{Temam_1969}
Roger Temam.
\newblock Sur l’approximation de la solution des équations de
  {Navier-Stokes} par la méthode des pas fractionnaires (ii).
\newblock \emph{Archive for rational mechanics and analysis}, 33\penalty0
  (5):\penalty0 377–385, 1969.

\bibitem[van~der Vorst(1992)]{bi-cgstab92}
H.~van~der Vorst.
\newblock Bi-{CGSTAB}: A fast and smoothly converging variant of bi-{CG} for
  the solution of nonsymmetric linear systems.
\newblock \emph{{SIAM} Journal on Scientific and Statistical Computing},
  13\penalty0 (2):\penalty0 631--644, 1992.

\bibitem[Weller et~al.(1998)Weller, Tabor, Jasak, and
  Fureby]{Weller_Tabor_Jasak_Fureby_1998}
H.~G. Weller, G.~Tabor, H.~Jasak, and C.~Fureby.
\newblock A tensorial approach to computational continuum mechanics using
  object-oriented techniques.
\newblock \emph{Computers in Physics}, 12\penalty0 (6):\penalty0 620–631, 11
  1998.

\bibitem[Zhang(2005)]{zhang_schur_2005}
Fuzhen Zhang, editor.
\newblock \emph{The Schur Complement and Its Applications}.
\newblock Numerical Methods and Algorithms. Springer {US}, 2005.
\newblock ISBN 978-0-387-24271-2.

\end{thebibliography}

\end{document}